# Construction of FuzzyFind Dictionary using Golay Coding Transformation for Searching Applications

Kamran Kowsari[†], Maryam Yammahi[†,] Nima Bari[†], Roman Vichr[*],
Faisal Alsaby[†], Simon Y. Berkovich[†]

[†] Department of Computer Science,
School of Engineering and Applied Sciences at
The George Washington University,
Washington DC, 20052

[*] Exprentis, Inc
Department of Data Mining and Engineering
Fairfax VA, 22030
Kowsari@gwu.edu or Berkov@gwu.edu

*Abstract*—searching through a large volume of data is very critical for companies, scientists, and searching engines applications due to time complexity and memory complexity. In this paper, a new technique of generating FuzzyFind Dictionary for text mining was introduced. We simply mapped the 23 bits of the English alphabet into a FuzzyFind Dictionary or more than 23 bits by using more FuzzyFind Dictionary, and reflecting the presence or absence of particular letters. This representation preserves closeness of word distortions in terms of closeness of the created binary vectors within Hamming distance of 2 deviations. This paper talks about the Golay Coding Transformation Hash Table and how it can be used on a FuzzyFind Dictionary as a new technology for using in searching through big data. This method is introduced by linear time complexity for generating the dictionary and constant time complexity to access the data and update by new data sets, also updating for new data sets is linear time depends on new data points. This technique is based on searching only for letters of English that each segment has 23 bits, and also we have more than 23-bit and also it could work with more segments as reference table.

## I. Introduction

The interaction between a search engine and database requires that the database structure to be consistent with a search engine to search quickly, easily, and efficiently. Golay Coding clustering technique which has faster time complexity in comparison with previous conventional methods such as K-means, spectral clustering and hierarchical clustering [8, 11] . Also the traditional method of clustering cannot cover fuzzy logic. This method is used for error correction, clustering, and other aspect in computer science. Our aim is to first modify how we utilize the Golay Code Clustering Hash Table (GCCHT), generates the Golay Code Transformation Hash Table (GCTHT), and finally creating the FuzzyFind Dictionary for searching thought Big Data. We use Golay Code (Golay Code Clustering Hash Table "GCCHT") because of its time complexity of this method and the fuzziness aspect, although a lot of research has been done on clustering. According to, research and projects of Dr. Arai K. the computational time for one data set by using Fuzzy C-means is around 80 ms in average, so therefore, we cannot use Fuzzy C-means for big data, and clustering is a method to find alike data point [10, 17].

## II. Related Work

When taking a look at the history of application development, we find out that many applications were programmed with a fuzzy logic component in low of classical and zero-one logics because the latter two were incapable of representing many datasets and solving scientific problems. Creating an effective fuzzy searching algorithms with a fast time complexity has proven difficult for researchers. Some groups have faced obstacles in their approach to Fuzzy Clustering. In 2009, the problem lie in the inability to search more than one keyword. The implementation of prefix queries for multiple keyword searching consisted of multiple algorithms such as ranking answer highlighting results and utilizing synonyms [15]. In 2011, some research group works on the fuzzy keyword searching on encrypted cloud storage data with small indices [13], the problem of that method is time complexity of searching which is completely dependent on length of keyword and edit distance. In this paper we focus on the one mathematical model that was introduced in the year of 1949 by Golay, Marcel JE for digital coding [23]. In between 1979 and 1981, NASA[1] in project of deep space missions developed this algorithms as error correction technique by using hamming distance [16, 19] and in some research project in year of 1969, the main challenges of computer science committee was working on this statistical method as radio communication for Gilbert burst-error-correcting [12, 21], but that binary Golay Code works with 24 bits which is not perfect Golay coding algorithms and the best Golay algorithms which can run in linear time complexity is 23 bits [18]. This method is an implementation of Golay transformation in conjunction with 23 bits and allows for error correction with two hashes utilizing the overlapping hash values [14]. In previous experiments, we concluded using the hash table is the most efficient method way for accessing data in constant time [13], Since 2014 we have been able to use the GCTHT in constructing a 23-bit meta-knowledge template for Big Data Discovery allowing for meta-feature extraction for clustering Structured and Unstructured Data (text-based and multimedia) [1, 2]. Yet the traditional use of GCTHT was hampering the ability to use fuzzy logic and we also realized the order of extracted meta-features was critical for Big Data clusterization. Re-arranging the order altered the results of the output ultimately requiring us to either use another algorithm to derive the best order or manually reorder to achieve best the best results [1, 2]. The other problems of that project is they use traditional way of GCTHT for their works and authors[1, 2] did not use the fuzzy logics of Golay Code, and another problem of these methods is the order of the features is critical for clustering the big data that means if we change the order of the features, the result will be changed in our output, so they need another algorithm for finding the best order of features or even they need to change the order of feature manually to find out best results. Intelligent Software Defined Storage is another example of usage of Golay Code Transformation Table [3].

---

[1] National Aeronautics and Space Administration (NASA)

### III. METHODOLOGY

The technique we use in this paper is divided to 3 main part; first, we need to create the Golay Code Clustering Hash Table (GCCHT) where we don't go through it because it's implemented before [7, 24] , The Golay Code Clustering Hash Table data structure uses hash by $2^{23}$ indices and each index could have six 12-bits labels or one 12-bit label where 86.47 percent of indices of GCCHT has six labels and the rest has only one label . After The GCCHT is generated, we need to generate the mapping hash table or Golay Code Transformation Hash Table [18]. Because GCTHT is generated by GCCHT it has same data structure we 86.47 percent by 15 23-bit value and the rest has one 23-bit value. The mapping of GCTHT is not efficient for searching because 13.53 percent of this table has only one 23-bit value and if we apply this algorithm for searching thought big data, absolutely we don't have accuracy. So we need to generate an efficient method of dictionary by using GCTHT. The generation of FuzzyFind Dictionary also has same data structure as GCTHT, but more than 98 percent of this hash table has 16 23-bit addresses and less than 2 percent has only one address.

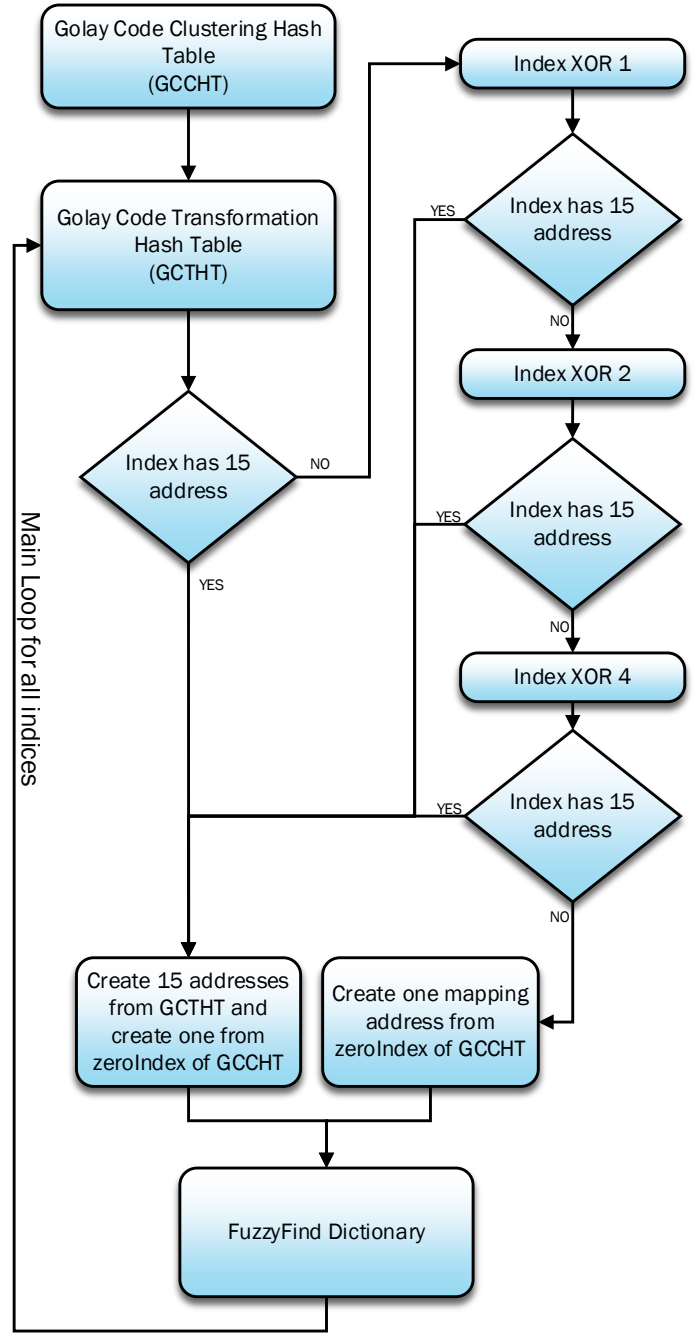

Figure 1 : Pipeline of generating FuzzyFind Dictionary using Golay Code Clustering Hash Table (GCCHT) and indicates the generating Golay Code Transformation Hash Table (GCTHT), and finally create FuzzyFind Dictionary, *"This algorithms is pre-processing algorithm for text manning and searching through big data"*

## B. Fuzzy Searching

Statistical data search engine, fuzzy Searching algorithms, when we can say that Fuzzy search algorithms are gained, modelling are designed by unstructured data from mathematical fuzziness modeling. The Fuzzy-Go search engine can thus automatically retrieve web pages that contain synonyms or terms similar to keywords, Figure 4 [13].

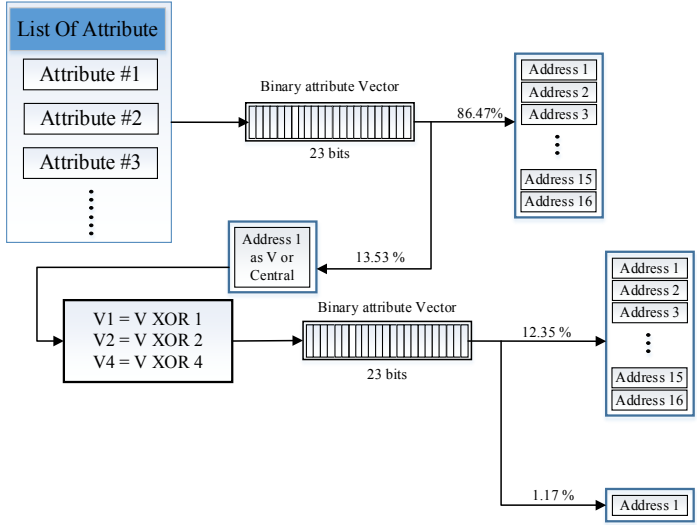

Figure 2 : General view of FuzzyFind Dictionary and indicates how improve the fuzziness logics from GCTHT [4, 7]

---

**Algorithm 1: Generating GCTHT by GCCHT**
1. **loop** $i = 0$ to 6
2.    **loop** $j = i+1$ to 6
3.       shift 12 bit of index one
4.       HashPair = shifted hash1 OR hash 2
5.       Shift one bit to right of HashPair
6.       Shifted HashPair XOR HashPair
7.    **end** of Loop 1
8. **end** of Loop 2

---

### A. Golay Code Transformation Hash Table

In the Big Data environment an explicit utilization of all available information is not feasible. Resolution of this situation requires formation of knowledge that would render a substantial part of data as "less interesting" [20]. Human reaction to the Big Data avalanche is bounded rationality - a decision-making process complying with cognition limitations and imposed deadlines. The ideology of bounded rationality leads to a computational model of the brain that goes beyond the traditional Turing algorithmic revealing unconsciousness as the basis for sophistication [6]. Golay Code Transformation Hash Table (GCTHT) is hash table with $2^{23}$ indices from 0 to $2^{23} - 1$ record and 15 addresses items that are created by Golay Code fuzzy clustering table [18]. The GCTHT is created by Golay Code Clustering Hash Table (GCCHT) that include from 0 to $2^{23}-1$ that contains 86.47 percent of them has six indices (labeled as fuzziness method) and 13.53 percent only has one indices (classical clustering, each data point belongs to one label) which means only 86.47 peents use real fuzzy logics clustering method and the rest following traditional clustering methods. GCTHT is created by GCCHT by following algorithms. GCTHT is created by all combination of labels.

$$\frac{6!}{4! \times 2!} = 15 \qquad (1)$$

$$|C| \sum_{k=0}^{e} \binom{n}{k} \leq 2^n \qquad (2)$$

$$2^{12} \sum_{k=0}^{n} \left( \binom{23}{0} + \binom{23}{1} + \binom{23}{2} + \binom{23}{3} \right) = 2^{23} \qquad (3)$$

---

Figure 3 : FuzzyFind Dictionary data Structure is presented in this figure by use Hash Table.

---

**Algorithm 2:** Generating Golay Code Clustering Hash Table
1.    **loop** $i = 0$ to $2^{23}$ //loop 1
2.       **loop** 1 to 24 //loop 2
3.          transform = (1 << i) & MASK_23; // Mask 23 bit all bits are 1
4.          codewordB = codeword XOR transform;
5.          recd = codewordB;
6.          recd = recd XOR decoding_table[get_syndrome(recd)];
7.          hash [i] = recd >> 11;
8.       **end** of loop 2
9.       Save Zero-Index (Zero index is needed for FuzzyFind Dictionary to create one address)
10.      **if** Hash elements have 2 different values **Then**
11.         Sort The Hash with 24 elements (It has 6 different values)
12.         Save in Golay Code hashes by 6 indices
13.      **end** of if
14.      **if** Hashes elements do not have 2 different values **then**
15.         Save NO-Value as first elements and save 1-index as Second Value
16. **end** of loop 1

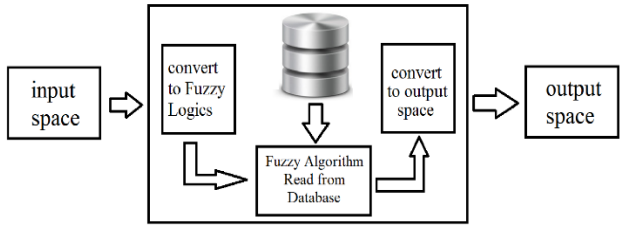

Figure 4 : This figure indicates how Fuzzy Search Engine works with input space and output space

*C. FuzzyFind Data Structure*

The FuzzyFind Search data structure works hash $2^{23}$ indices (0 to 8,388,607) and each indexes hash 16 different address as 23 bits from (0 to 8,388,607). The FuzzyFind Dictionary Data Structure contains by 16 addresses, but as regarding to 0.16 percent of this tables has only one address in mapping hash table. Construction of a dictionary is one of the basic undertakings in data structure developments. A dictionary is in essence a table that takes a key as an entry and if the key is present in the table returns some information associated with this key [20].

*D. FuzzyFind Dictionary*

Nowadays, fuzzy logic is used by a variety of search engines and cloud computing. Keyword based search methods allow us to select the retrieve files or words and has also been widely applied in plaintext search scenarios [22], such as Google or Bing search engine. The hash transformation utilizes the Golay Code. the decoding procedure which takes neighborhood spheres of radius 1surrounding the 23-bit binary vectors and yields hash codes as 12-bit keys: six hashes in 86.47%[7] of the cases(let's call it Case A) and one hash in 13.53% of the cases(let's call it Case B). In Case A, we apply Golay into a 23-bit vector to obtain 6 indices at 86.47% and further apply a transformation of the six 12-bit indices into 15 pairs. For Case B we apply Golay Coding into a 23-bit vector as we did with Case A, but will obtain 1-index which contains 13.53% of whole the GCTHT dividing it into two different categories. We apply Hamming Distances (HD) of 1 and 2 by yield and follow Algorithm 3. The Hamming distance between two codewords is equal to the number of bits in which they differ. It shows that if the error code and other code wants to become a few bits that must be changed to make this conversion is done without the error of the systems, brought to account. Hamming distance between two strings is equal to the length of information theory where the corresponding symbols are different. In other words, the minimum number of alternatives that will change one string to another string, or the number of errors that will convert a string to another string [22].

**Algorithm 3:** Generating FuzzyFind Dictionary using GCTHAT

1. **loop** $i = 0$ to $2^{23} - 1$
2.    **if** it does not have 15 mapping addresses **Then**
3.       index XOR 1
4.       **if** it does not have 15 mapping addresses **Then**
5.          Index XOR 2
6.          **if** it does not have 15 mapping addresses **Then**
7.             Index XOR 4
8.             **if** it does not have 15 mapping addresses **Then**
9.                Create one addresses with one label
10.          **else** of if 3
11.             Create one address with one label and 15 addresses from mapping
12.       **else** of if 2
13.          Create one address with one label and 15 addresses from mapping
14.    **else** of if 1
15.       Create one address with one label and 15 addresses from mapping
16. **end** of Loop

1000 → 00000000000001111101000
480 → 00000000000000111100000

| 1000 → | | 480 → | |
|---|---|---|---|
| [0] | 5244416 | [0] | 256043 |
| [1] | 202860 | [1] | 43 |
| [2] | 202947 | [2] | 108 |
| [3] | 203166 | [3] | 199 |
| [4] | 204288 | [4] | 1920 |
| [5] | 202496 | [5] | 3840 |
| [6] | 442563 | [6] | 176236 |
| [7] | 442782 | [7] | 176327 |
| [8] | 443904 | [8] | 178048 |
| [9] | 446208 | [9] | 179968 |
| [10] | 799134 | [10] | 442567 |
| [11] | 800256 | [11] | 444288 |
| [12] | 802560 | [12] | 446208 |
| [13] | 1697280 | [13] | 819072 |
| [14] | 1699584 | [14] | 816896 |
| [15] | 6295296 | [15] | 7868160 |

Figure 5 : Sample result of FuzzyFind Dictionary and shows that two indices with hamming distance of less or equal than two has at least one same mapping index.

## IV. RESULTS

For testing FuzzyFind Dictionary we needs to test all cases which means Case A and B with 16 addresses and Case C with only one address. In our test algorithm we need to select one or more random word and create and look at the addresses of the theses word, after that find all hamming distance of this index within less and equal to 2 which means HD with 0, 1, and 2. As regarding to 23 bits indices we have 253 results for each indices with HD of 2 and 23 results for HD of 1 and only one index for HD is equal to 0, so the test algorithm needs to consider to 277 results [7]. After finding this indices, the algorithms must find at least one same address in the FuzzyFind Dictionary, all of them should have same address with our index which we choose. If the test algorithm find same address for all 277 indices, our FuzzyFind Dictionary is working, but even one of the test indices does not have same address with index we choose, The FuzzyFind Dictionary is not working.

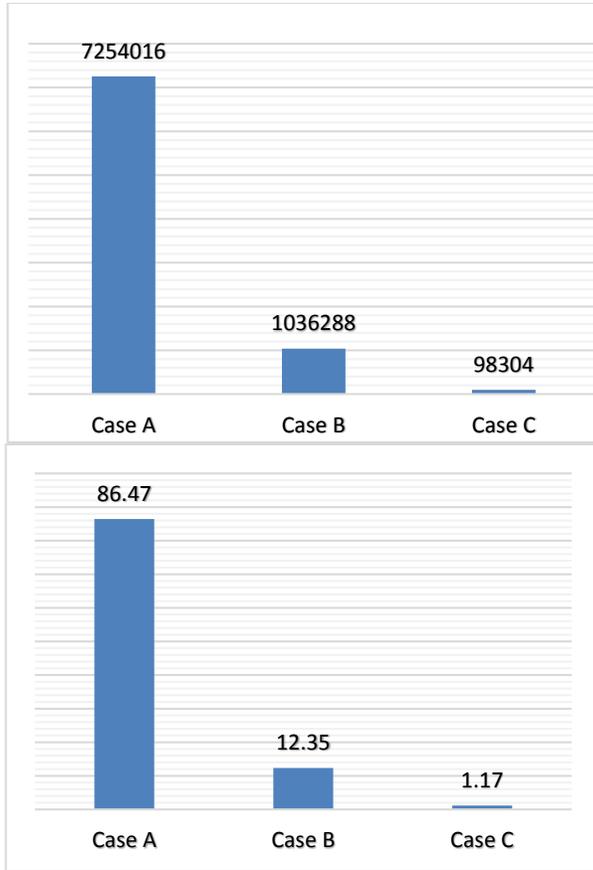

Figure 6 : Number of Case A, Case B and Case C where Case A is the number of indices which has 15 addresses into GCTHT, Case B is number of indices which has 1 indices but we can find 15 addresses by using Algorithm 1, and Case C indicates the number of indices we could not find any 15 addresses by using algorithm and only we create 1 addresses from Zeroindex of GCCHT, Chart B indicates Percentage of each group.

V. DISCUSSION

The result of Case A is about 86.47 percent which has 16 different addresses and Case B that has one 12 bit hash but only can find 16 addresses by using V1, V2 and V4 (one 12-bit hash XOR with 1, 2 and 4) and algorithm can find 16 addresses is around 12.35 percent (Figure 2), so by adding this two percent we can find out 98.83 percent of our FuzzyFind Dictionary has 16 addresses and less than 1.16 percent only has 1 address. In our implementation, called FuzzyFind Dictionary, we simply mapped the 26 letters of the English alphabet into 23 bits, reflecting the presence or absence of particular letters. This representation preserves closeness of word distortions in terms of closeness of the created binary vectors. Within Hamming distance 2 deviation. A hash transformation using the Golay code decoding procedure is applied to neighborhood spheres of radius 1 surrounding 23-bit binary vector [7]. We assume that you know already about the Golay Coding transformation. There are three basic provisions in realization of fuzzy retrieval with the suggested data structure (FuzzyFind Dictionary):

(1) Attributes of information items have to be mapped to a binary vector in such a way that closeness in attribute discrepancies is translated into closeness of binary vectors in Hamming's metric,

(2) The format of the fault-tolerant indexing based on this scheme imposes limitations on the length of the binary vector, in the case of Golay code the length is 23, and

(3) The retrieved binary vectors cannot deviate from the search vector more than a relatively small value of the Hamming distance, typically by 2.

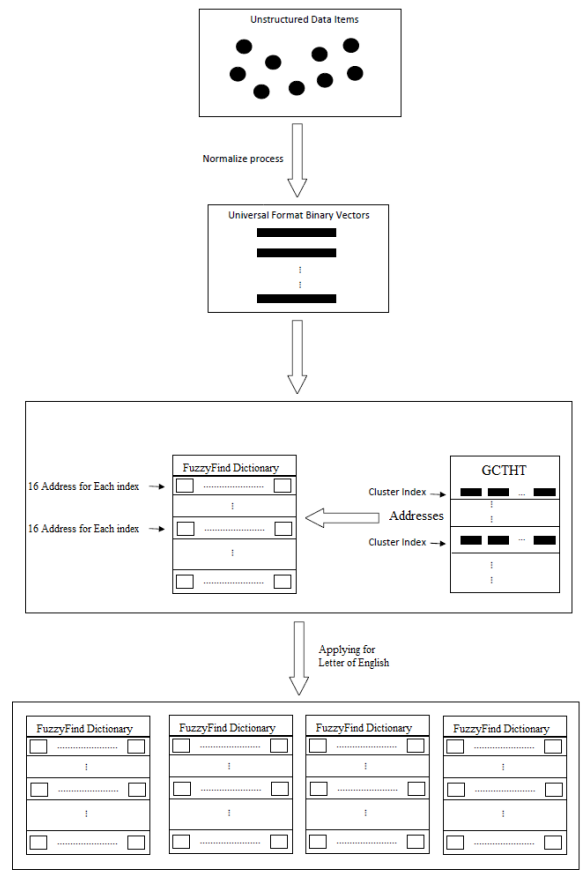

Figure 7 : This figure represent the fact that how we can use FuzzyFind Dictionary for more than 23 bits

## VI. Conclusion and Future Work

Perhaps, all of the information provided in this paper research, might reflect the fact that generating FuzzyFind Dictionary by using Golay coding transformation Hash Tables is one of the efficient method of creating Fuzzy Dictionary for searching through big data. The Golay Coding Transformation Hash Table (GCTHT) and Golay Coding Address Tables (GCAT) works with linear time complexity and after creating GCTHT just we need to apply the into FFD algorithms by linear time complexity by following steps: First GCTHT is generated into hash table, so the algorithms has access by O(1) time complexity; second ,Generating FFD for all 2 power 23 needs linear time complexity as regarding to: Case A: They have six unique 12-bit indices and use 15 addresses of from GCAT and create another address with zero-indexes hash table. Case B: They have one 12-bit index, and one address, so we could find 15 addresses by the nearest data set by XOR 1, 2 or 4 and find 15 addresses by yield. Case C: these part is same with Case B, but FFD has only one address because of nearest data point which XOR with 1, 2 and 4 also has one address. This method has good time complexity for generating FuzzyFind Dictionary which is Linear, O(n) and assess to the FFD is constant time complexity O(1) because FFD used hash table. In this paper we show that a FuzzyFind Dictionary improved percentage of indices with sixteen addresses by 98.83 from GCTHT with 86.47 percent with 15 addresses. This FuzzyFind Dictionary can be used into search engine and also this method can be used in error correction of miss typing and sudden interruption of communication, loss, or lack of landmarks, fields containing Null, abbreviations unusual or abnormal fields are experiencing any reason. We have plan to use FuzzyFind Dictionary for indexing by supervised learning method and using historical data points. We have plan to distribute the source code and improve it for searching application in near future. Also this algorithms can be useful for mining gene sequences because the main challenge of bioinformatics, biology, and biological scientist is RNA-seq mining datasets [5]. We have plan to implement Fuzzy logics for DNA and RNA analyses which can be useful for diagnosis Tumor [9] as FuzzyFind logics.

## VII. Acknowledgment

We would like to sincerely thank Professor Simon Y. Berkovich for his guidance and a special thanks to our all faculty and staff of Department of Computer Science of The school of Engineering and applied Science at the George Washington University for their support.

IX. AUTHOR'S PROFILE

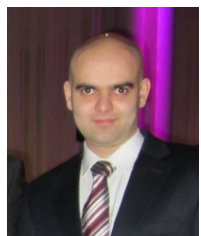
**Kamran Kowsari,** He is a Ph.D Candidate in Computer Science at The George Washington University Washington DC USA. He received his Master of Science in Computer Science, Data Mining and Machine Learning at The George Washington University Washington DC USA. He has more than 6 year's experiences in software development, system and database engineering experience, and research. His experience includes numerous projects and academic projects. He is interested and has experience in Machine learning, mathematical modeling, algorithms and data structure, Real-time rendering use machine learning, and data manning, Computer Graphics and visualization, volumetric Rendering .

**Maryam Yammahi,** She received her Ph.D degree in Computer Science at The George Washington University, her focus area is Data manning and Machine Learning, Fuzziness Searching, mathematical model.

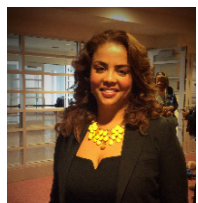
**Nima Bari**, She received her Ph.D degree in Computer Science at The George Washington University, her focus area is Data manning and Machine Learning, Fuzziness clustering, meta-knowledge and knowledge discovery.

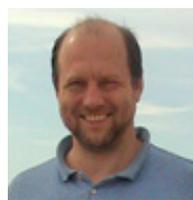
**Roman Vichr**, He received his PhD. degree in Material Engineering from the Institute of Chemical Technology of Prague in 1992. He has more than 20 year's experiences in international software development, system and database engineering experience. His experience includes numerous projects for Fortune 100 companies and government bodies at the federal and state level.

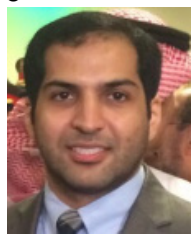
**Faisal Alsaby**, He is currently a Ph.D candidate at the GWU majoring in Computer Science. He received an MS degree in Computer Science from the George Washington University, Washington, DC, and USA in 2012. His research interests are big data clustering algorithms, machine learning.

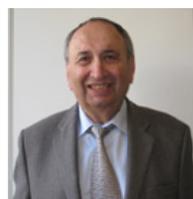
**Simon Y. Berkovich**, He received his Ph.D in Computer Science, 1964, Institute of Precision Mechanics and Computer Technology, USSR Academy of Sciences, his MS in Applied Physics, 1960, Moscow Physico-Technological Institute, He is Professor and Faculty  at School of Engineering and applied Science at The George Washington University. His major area of research is information retrieval, computer organization, and mathematical modeling. He has more than 100 technical publications and 30 patent.